\documentclass{sig-alt-full}
\usepackage{graphicx}
\usepackage{cite}
\usepackage{url}
\begin{document}

\conferenceinfo{SCG'05,} {June 6--8, 2005, Pisa, Italy.}
\CopyrightYear{2005}
\crdata{1-58113-991-8/05/0006}

\newtheorem{lemma}{Lemma}
\newtheorem{theorem}{Theorem}
\newtheorem{definition}{Definition}

\title{Minimum Dilation Stars}
\numberofauthors{1}
\author{
\alignauthor David Eppstein and Kevin A. Wortman\\
   \affaddr{Department of Computer Science}\\
   \affaddr{University of California, Irvine}\\
   \email{\{eppstein, kwortman\}@ics.uci.edu}\\
}

\maketitle

\begin{abstract}
The dilation of a Euclidean graph is defined as the ratio of distance
in the graph divided by distance in $\mathbb{R}^d$.  In this paper we
consider the problem of positioning the root of a star such that the
dilation of the resulting star is minimal.  We present a deterministic
$O(n \log n)$-time algorithm for evaluating the dilation of a given
star; a randomized $O(n \log n)$ expected-time algorithm for finding
an optimal center in $\mathbb{R}^d$; and for the case $d=2$, a randomized
$O(n \, 2^{\alpha(n)} \log^2 n)$ expected-time algorithm for finding
an optimal center among the input points.
\end{abstract}



\keywords{Facility location, quasiconvex programming, dilation,
stretch factor, Euclidean graphs}

\section{Introduction}

A \emph{star} is a graph with exactly one internal vertex, called its
\emph{center}; it has edges from the center to every external vertex,
and no other edges.  The \emph{dilation} between any pair of vertices
$a$ and $b$ in a Euclidean graph is defined \cite{Epp-HCG-00} as the
cost of the shortest path from $a$ to $b$, divided by the Euclidean
distance $|ab|$.  The dilation of a graph is defined as the maximum
dilation over all pairs of vertices.  We consider the following
problem: for any fixed $d \geq 2$, given a set $V \subset \mathbb{R}^d$ of
$n$ points, construct a star with center $c \in \mathbb{R}^d$ and leaves
$V$ such that the star has minimal dilation.  We present algorithms
for the case when $c$ is constrained to be one of the input points, as
well as the unconstrained case when $c$ may be any point in $\mathbb{R}^d$.

We present algorithms to evaluate the dilation of a given star in $O(n
\log n)$ time; find the optimal center $c \in \mathbb{R}^d$ for a given set
of external vertices in $O(n \log n)$ expected time; and for a set of
vertices $V \subset \mathbb{R}^2$, select an optimal center $c \in V$ in
expected $O(n \, 2^{\alpha(n)} \log^2 n)$ time.  The evaluation
algorithm's task is to find the single pair of points that define the
graph's dilation.  It works by identifying $O(n)$ pairs of points with
the potential to be that pair in $O(n \log n)$ time, and evaluating
the dilation for each of those $O(n)$ point pairs.  Our algorithm for
the unconstrained optimization algorithm formulates the problem as a
quasiconvex program, and uses the evaluation algorithm as a component
of Timothy Chan's randomized optimization framework \cite{982853} to
arrive at the solution in $O(n \log n)$ expected time.  The algorithm
for the constrained case works by repeatedly selecting a random
vertex, evaluating the dilation that would result from using that
vertex as the center, computing the region $R$ of space that could
contain a center yielding an even lower dilation, and discarding all
the vertices outside $R$.  This procedure is iterated an expected
$O(\log n)$ times, and computing a description of $R$ takes $O(n \,
2^{\alpha(n)} \log n)$ time, resulting in an $O(n \, 2^{\alpha(n)}
\log^2 n)$ expected-time algorithm.

Finding a minimum dilation star can be viewed as an instance of the
classic facility location problem \cite{location-analysis}: given a
set of demand sites and supply sites, find a center minimizing a
particular objective function.  In our case the objective function is
defined as the maximum dilation between any pair of points.  This
formulation could e.g. represent the problem of deciding where to
position an airline hub, such that flights with layovers are not
unreasonably longer than direct flights.

Narasimhan and Smid \cite{586967} considered the related problem of
computing dilation approximately.  They present algorithms which can
compute an $\varepsilon$-approximation of the dilation of a path in
$O(n \log n)$ time, a cycle in $O(n \log n)$ time, and a tree in $O(n
\log^2 n)$ time, for any fixed $d \geq 1$.  Agarwal et
al. \cite{akks-cdpc-02} considered the problem of computing the
dilation of a polygonal curve.  They present algorithms to compute the
exact dilation of a polygonal curve in $\mathbb{R}^2$ in $O(n \log n)$
expected time, or in $\mathbb{R}^3$ in subquadratic time.  Langerman
et al. \cite{696169} presented algorithms which compute the dilation
of a planar polygonal curve in expected $O(n \log n)$ time, a planar
tree in $O(n \log^2 n)$ time, or a planar cycle in $O(n^{\frac{3}{2}}
\log n)$ time.  The evaluation algorithm presented in this \pagebreak
paper computes the exact dilation of a star in $O(n \log n)$ time for
any fixed $d \geq 2$.

\section{The evaluation problem}

\label{evaluation}

Formally, the evaluation problem is as follows: given a set $V$ of $n$
points in $\mathbb{R}^d$ and a point $c \in \mathbb{R}^d$, compute the
dilation of the Euclidean star with center $c$ and leaves $V$.

For any pair of vertices $a$ and $b$, a star contains exactly one path
from $a$ to $b$: the path $\langle a, c, b \rangle$.  So the dilation
between $a$ and $b$ is
\[ \Delta_c(a,b) = \frac{|ac| + |cb|}{|ab|} \]
where $|ac|$ designates the Euclidean distance from $a$ to $c$.

The dilation of the star is
\[ \Delta_c = \max_{a, b \in V} {\Delta_c(a,b)} \enspace . \]

Define $\hat{a}$ to be one of the points in $V$ for which $\Delta_c(a,
\hat{a})$ is greatest (there may be more than one pair of points with
dilation $\Delta_c$).  As noted in the introduction, one approach to
solving this problem is to reason that for any $a \in V$, $\hat{a}$
could be any of the $n-1$ points in $V \setminus\{a\}$.  Then
$\Delta_c$ may be computed by an algorithm that evaluates
$\Delta_c(a,b)$ for all $O(n^2)$ pairs of points.  As we will see,
this approach is overly conservative.  In this section we present an
algorithm that identifies $O(n)$ candidate pairs of points, such that
the pair $(a, \hat{a})$ for which $\Delta_c=\Delta_c(a,\hat{a})$ is
guaranteed to be one of the candidates.  The algorithm takes $O(n \log
n)$ time to generate the candidate list, and computing the largest
dilation among $O(n)$ pairs of points takes $O(n)$ time.  The point
pairs are identified using two techniques, each generating $O(n)$
pairs.  One technique generates a list of $O(n)$ pairs of points
guaranteed to contain $(a,\hat{a})$ when $\Delta_c$ is high; the other
generates $O(n)$ pairs guaranteed to contain $(a,\hat{a})$ when
$\Delta_c$ is low.

\subsection{The high dilation case}

\label{high}

One of our heuristics for finding $\hat{a}$ is to identify $a$'s $k$
nearest neighbors, for a constant $k \in O(1)$.  This heuristic is
guaranteed to identify $\hat{a}$ when $\Delta_c=\Delta_c(a,\hat{a})$
is high; formally, we will prove that if $\Delta_c \geq \Gamma$ for a
constant $\Gamma>3$, then $\hat{a}$ is one of $a$'s $k$ nearest
neighbors, for a constant $k$.  The $k$ nearest neighbors of every $a
\in V$ may be reported in $O(k n \log n)$ time using the algorithm of
Vaidya \cite{70532}, so the process of identifying these $O(n)$ point
pairs takes $O(n \log n)$ time.

\begin{lemma}
\label{c-out-of-kappa}
Suppose $a \in V$ and $\Delta_c = \Delta_c(a,\hat{a}) \geq \Gamma$ for
a constant $\Gamma>3$.  Then $|ac| \geq \phi |a \hat{a}|$ for a
constant $\phi>1$ that depends only on $\Gamma$.
\end{lemma}

\begin{proof}

By definition,

\[ \Gamma  \leq  \Delta_c = \frac{|ac|+|c\hat{a}|}{|a\hat{a}|} 
 \leq  \frac{|ac|+(|ac|+|a \hat{a}|)}{|a \hat{a}|} 
 \leq  1 + 2 \frac{|ac|}{|a \hat{a}|} \enspace , \]
so
\[ |ac| \geq \frac{\Gamma - 1}{2} |a \hat{a}| \enspace .\]
Therefore if $\Gamma > 3$, $\phi=\frac{\Gamma-1}{2} >
1$. 

\hspace{7.55cm} \end{proof}

\begin{lemma}
\label{uv-spaced}
Suppose that $\Delta_c = \Delta_c(a,\hat{a}) \geq \Gamma$ for a
constant $\Gamma>3$, let $\kappa$ be the $d$-sphere centered on $a$
with radius $|a \hat{a}|$, and $u, v \in V$ be two input points inside
$\kappa$.  Then there is a constant $\gamma>0$ that depends only on
$\Gamma$, such that $|uv| \geq \gamma |a \hat{a}|$.
\end{lemma}

\begin{proof}
By Lemma~\ref{c-out-of-kappa}, $|ac|-|a \hat{a}| \geq (1-1/\phi)|ac|$, so

\begin{eqnarray*}
\frac{3 |ac|}{|a \hat{a}|} & \geq & \frac{|ac|+|\hat{a}c|}{|a\hat{a}|}
= \Delta_c \geq \Delta_c(u,v) \\ & \geq & \frac{|uc|+|cv|}{|uv|} \geq
\frac{2(|ac|-|a \hat{a}|)}{|uv|} \\ & \geq &
\frac{2(1-1/\phi)|ac|}{|uv|} \enspace .
\end{eqnarray*}

It follows that $|uv| \geq
\frac{2}{3}(1-1/\phi)|a \hat{a}|$. So $|uv| \geq \gamma |a \hat{a}|$
for $\gamma=\frac{2}{3}(1-1/\phi)$; $\phi>1$ so $\gamma>0$. 

\hspace{7.55cm} \end{proof}

\begin{lemma}
\label{high-n-points}
If $\Delta_c \geq \Gamma$ and $\Delta_c = \Delta_c(a,\hat{a})$, then
$\hat{a}$ is one of $a$'s $k$ nearest neighbors, for a constant $k$
depending only on $\Gamma$ and the dimension.
\end{lemma}

\begin{figure}
\centering
\scalebox{.6}{\includegraphics[bb=0cm 0cm 8cm 8cm]{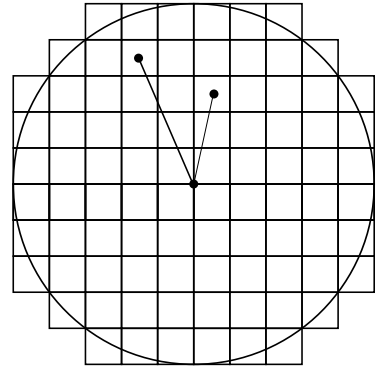}}
\caption{$\kappa$ partitioned into small $d$-cubes for $d=2$.}
\label{fig:squares}
\end{figure}

\begin{proof}

As shown in Lemma~\ref{uv-spaced}, the distance between any $u$ and
$v$ is proportional to the radius $r_\kappa = |a \hat{a}|$ of
$\kappa$. Let $\sigma = \gamma/\sqrt{d}$ and partition $\kappa$ into
small $d$-cubes with sides of length $\sigma r_\kappa$ (see
Figure~\ref{fig:squares}).  Under such an arrangement no cube contains
more than one input point.  $\kappa$ can contain no more than $\lceil
(\frac{ 2 r_\kappa }{\sigma r_\kappa})^d \rceil = \lceil
(\frac{2}{\sigma})^d \rceil$ such cubes, which is constant in $n$.  So
we may conclude that if $\Delta_c=\Delta_c(a,\hat{a})$ then $\hat{a}$
is one of $a$'s $k$ nearest neighbors, for a constant $k=\lceil
(\frac{ 2 }{\sigma})^d \rceil$. 

\hspace{7.55cm} \end{proof}

\subsection{The low dilation case}

\label{low}

Our other heuristic for finding $\hat{a}$ is to identify points whose
distance from $c$ is similar to that of $a$.  This may be done by
sorting $V$ by distance from $c$, then for each $a$, identifying the
points whose indices in the sorted sequence are within $l$ of $a$'s,
for a constant $l$.  We will show that this heuristic is guaranteed to
identify $\hat{a}$ when $\Delta_c=\Delta_c(a,\hat{a})$ is low, i.e.,
that if $\Delta_c \leq \Gamma$ for a constant $\Gamma$, and $V_S$ is
the set $V$ sorted by distance from $c$, then for any $v_i \in V_S$,
if $v_j = \hat{v_i}$ and $\Delta_c=\Delta_c(v_i,v_j)$ then $|i-j| \leq
l$, for a constant $l$.  We prove this by showing that space may be
partitioned into $d$-dimensional annuli with exponentially growing
radii, all centered on $c$, such that each annulus contains $O(1)$
points from $V$, and for any $a \in V$, $\hat{a}$ lies within one of
$O(1)$ adjacent annuli.  We define an annulus $A_{c,i}$ in terms of
its inner and outer radii, $\rho^i$ and $\rho^{i+1}$, respectively,

\[ A_{c,i} = \{ p \in \mathbb{R}^d \enspace | \enspace \rho^i \leq |pc| \leq \rho^{i+1} \} \]
where the constant $\rho$ is defined as
\[ \rho > \sqrt{\frac{\Delta_c+1}{\Delta_c-1}} \enspace .\]

\begin{lemma}
\label{constant-annuli}
If $a \in A_{c,i}$ and $\hat{a} \in A_{c,i+k}$, then $k \leq 2$.
\end{lemma}

\begin{proof}
Suppose for the sake of contradiction that $k \geq 3$.  We now
establish an upper bound on $\Delta_c(a,\hat{a})$ by positioning $a$
and $\hat{a}$ such that dilation is maximized.  This happens when $a$
and $\hat{a}$ are colinear and as close as possible.  We may assume
without loss of generality $|ac| \leq |\hat{a}c|$; so
$|ac|=\rho^{i+1}$ and $|\hat{a}c|=\rho^{i+3}$, hence

\begin{eqnarray*}
\Delta_c(a,\hat{a}) & = & \frac{\rho^{i+1} +
\rho^{i+3}}{\rho^{i+3}-\rho^{i+1}} \\ &=& \frac{1+\rho^2}{\rho^2-1} \\
&<&
\frac{1+\frac{\Delta_c+1}{\Delta_c-1}}{\frac{\Delta_c+1}{\Delta_c-1}-1}
= \frac{(\Delta_c-1) + \Delta_c +1}{\Delta_c+1 - (\Delta_c - 1)} =
\Delta_c
\end{eqnarray*}
which is a contradiction; so the assumption $k \geq 3$ must be
false. 

\hspace{7.55cm} \end{proof}

\begin{lemma}
\label{lower-angle-bound}
There exists a function $\theta_{MIN}$ such that if $a$ and $b$ lie in
the same annulus, $\angle acb \geq \theta_{MIN}(\Delta_c)$, and a
value $\Delta_0$ such that for all $\Delta_c < \Delta_0$,
$\theta_{MIN}(\Delta_c) > \pi/2$.
\end{lemma}

\begin{proof}
We wish to establish a lower bound on $\angle acb$, and so we consider
the choice of $a$ and $b$ that makes $\angle acb$ as acute as
possible.  For a fixed $\Delta_c(a,b)$, we can do this by making $|
|ac|-|cb| |$ as large as possible.  Without loss of generality, assume
$|cb|=\rho|ac|$; then
\[ \Delta_c \geq \frac{|ac|+\rho|ac|}{|ab|} \enspace .\]
Let $\theta=\angle acb$; then by the law of cosines,
\begin{eqnarray*}
\Delta_c & \geq & \frac{|ac| + \rho |ac|}{\sqrt{|ac|^2 + \rho^2 |ac|^2
- 2 \rho |ac|^2 \cos \theta}} \\ &\geq& \frac{1+\rho}{\sqrt{1+\rho^2-
2 \rho \cos \theta}} \enspace .
\end{eqnarray*}
Solving for $\theta$,
\[ \theta \geq \cos^{-1} ( \frac{\rho}{2} + \frac{1}{2 \rho} - \frac{(1+\rho)^2}{2 \rho \Delta_c^2} ) \enspace ,\]
so we let $\theta_{MIN}(\Delta_c) = \cos^{-1} ( \frac{\rho}{2} +
\frac{1}{2 \rho} - \frac{(1+\rho)^2}{2 \rho \Delta_c^2} )$.  The
result follows from the convergence of the limit
\[ \lim_{\Delta_c \rightarrow 1^{+}} \theta_{MIN}(\Delta_c) = \pi  \enspace .\] 

\hspace{7.55cm} \end{proof}

\begin{lemma}
\label{const-points-in-annulus}
Each $A_{c,i}$ contains $O(1)$ points from $V$.
\end{lemma}

\begin{proof}
Lemma~\ref{lower-angle-bound} shows that if $a$ and $b$ are in the
same annulus and $\theta = \angle acb$, then
\[ \theta \geq \theta_0 = \min_{\Delta_0 \leq \Delta_c \leq \Gamma} \theta_{MIN}(\Delta_c) \enspace . \]
Thus each annulus contains $O(\theta_0^{1-d})$ points; $\theta_0$ and
$d$ are constant with respect to $n$, so this quantity is as
well. 

\hspace{7.55cm} \end{proof}

\begin{lemma}
\label{low-n-points}
If $\Delta_c \leq \Gamma$, $V_S$ is the set $V$ sorted by distance
from $c$, $v_s$ and $v_t \in V_S$, and $\Delta_c=\Delta_c(v_s,v_t)$,
then $|s-t|\leq l$ for a constant $l$ depending only on $\Gamma$ and
$d$.
\end{lemma}

\begin{proof}
Lemma~\ref{constant-annuli} shows that if $a \in A_{c,i}$ and $\hat{a}
\in A_{c,j}$, then $|i-j|\leq 2$.  The points in contiguous annuli
will be contiguous in $V_S$.  So $\hat{a}$ must be within $l$ ranks of
$a$ in $V_S$, where $l$ is the number of points that may lie
in two annuli.  As shown in Lemma~\ref{const-points-in-annulus}, $l$
is constant with respect to $n$. 

\hspace{7.55cm} \end{proof}

\subsection{The evaluation algorithm}

\begin{theorem}
Given a set $V \subset \mathbb{R}^d$ of $n$ points and center $c \in
\mathbb{R}^d$, it is possible to compute the dilation $\Delta_c$ for
the star with center $c$ and leaves $V$ in $O(n \log n)$ time,
provided $d$ is constant.
\end{theorem}

\begin{proof}
We generate a list of $O(n)$ pairs of points as follows.  Compute the
$k$ nearest neighbors for the set $V$ in $O(n \log n)$ time, for the
constant $k$ described in Subsection~\ref{high}.  Then for each $a \in
V$, append the point pairs consisting of $a$ paired with each of $a$'s
$k$ nearest neighbors to the list.  Next, create $V_S$ by sorting the
points in $V$ by distance from $c$ in $O(n \log n)$ time.  For each
$v_i \in V_S$, add the pair $(v_i, v_j)$ to the list, for every $j$
such that $|i-j|\leq l$, for the constant $l$ described in
Subsection~\ref{low}.  The resulting list has $(kn+ln) \in O(n)$
elements.  As shown in Lemmas~\ref{high-n-points}
and~\ref{low-n-points}, the pair defining the dilation of the star,
$(a,\hat{a})$ such that $\Delta_c=\Delta_c(a,\hat{a})$, is certain to
be present in the list.  So we can compute $\Delta_c$ by evaluating
the dilation of each of the $O(n)$ pairs of points, and returning the
maximum. 

\hspace{7.55cm} \end{proof}

\section{The unconstrained \\ optimization problem}

\label{unconstrained}

We now turn to the following problem: given a set of $n$ points $V
\subset \mathbb{R}^d$, find a point $c$, not necessarily belonging to $V$,
such that the star with center $c$ and leaves $V$ has minimal
dilation.

\subsection{Reduction to a quasiconvex program on $O(n^2)$ contraints}

A function $f: \mathbb{R}^d \mapsto \mathbb{R}$ is \emph{quasiconvex}
when its level sets $f^{\leq \lambda}=\{x \in \mathbb{R}^d \enspace |
\enspace f(x) \leq \lambda \}$ are convex.  Quasiconvex programming
\cite{eppstein-quasiconvex} is a generalization of linear programming:
a \emph{quasiconvex program} is a finite set $S$ of quasiconvex
functions (constraints); the solution to a quasiconvex program is the
input $x \in \mathbb{R}^d$ minimizing the objective function $f_{S} =
\max_{f_i \in S} f_i(x)$.  Amenta et al. showed
\cite{AmeBerEpp-Algs-99} that if a solution to a constant-sized subset
of $S$ may be solved in $O(1)$ time, then the program may be solved in
$O(|S|)$ expected time.

\begin{lemma}
\label{dilation-qcp}
The star center $c$ minimizing $\Delta_c$ may be computed by solving a
quasiconvex program with $O(n^2)$ constraints.
\end{lemma}

\begin{figure}
\centering
\scalebox{.7}{\includegraphics[bb=0in 0in 2.5in 1.8in]{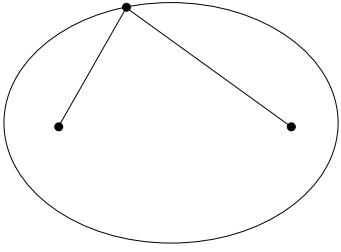}}
\caption{An example level set $f_{i,j}^{\leq \lambda}$.}
\label{fig:ellipse}
\end{figure}

\begin{proof}
We introduce one contraint $f_{i,j}: \mathbb{R}^d
\mapsto \mathbb{R}$ for each pair of points $v_i, v_j \in V$:
\[ f_{i,j}(x) = \frac{|v_i x|+|x v_j|}{|v_i v_j|} \enspace . \]
Hence the solution $x$ is precisely the point $c$ for which the
dilation is minimized for all pairs of input points.

The denominator of $f_{i,j}$ is the distance between $v_i$ and $v_j$,
which is constant with respect to $x$.  So $f_{i,j}(x)$ is
proportional to the sum of the distances of $x$ from $v_i$ and $v_j$,
or equivalently each $f_{i,j}$ defines an ellipsoid with foci $v_i$
and $v_j$.  So any level set $f_{i,j}^{\leq \lambda}$ is elliptical
(see Figure~\ref{fig:ellipse}), and hence convex; so each $f_{i,j}$ is
quasiconvex. 

\hspace{7.55cm} \end{proof}

\subsection{The optimization algorithm}

In this section we show how to solve the unconstrained optimization
problem in $O(n \log n)$ expected time.  We adapt Timothy Chan's
randomized optimization technique \cite{982853} to our problem.  Chan
states his result in terms of LP-type problems, but following
\cite{eppstein-quasiconvex}, we rephrase it in terms of quasiconvex
programming:

\begin{lemma}[Chan]
Let $\cal Q$ be a space of quasiconvex functions,
$\cal P$ be a space of input values,
and $f:2^{\cal P}\mapsto 2^{\cal Q}$ map sets of input values to sets of 
functions in $\cal Q$.
Further, suppose that $\cal P$, $f$, and $\cal S$ satisfy the following 
properties:
\begin{itemize}
\item There exists a constant-time subroutine for solving quasiconvex 
programs
of the form $f(B)$ for any $B\subset\cal P$ with $|B| \in O(1)$.
\item There exists a decision algorithm
that takes as input a set $P\subset\cal P$ and a pair $(\lambda,\bar x)$, 
and returns yes
if and only if $q(\bar x)\le\lambda$ for all $q\in f(P)$.
The running time of the decision algorithm is bounded by $D(|P|)$, where
there exists a constant $\epsilon>0$ such that $D(n)/n^{\epsilon}$ is 
monotone increasing.
\item There are constants $s$ and $r$ such that, for any input set 
$P\subset\cal P$, we can find in time at most $D(|P|)$ a collection of sets 
$P_i$,
$0\le i<r$, each of size at most $s |P|$, for which
$f(P)=\cup_i f(P_i)$.
\end{itemize}
Then for any $P\subset\cal P$ we can solve the quasiconvex program
$f(P)$, where $|P|=n$, in randomized expected time $O(D(n))$.
\end{lemma}

We now show how to apply this result to the problem at hand.

\begin{theorem}
\label{expected-time}
Given a set $V$ of $n$ points, it is possible to compute a point $c$
that admits the minimal-dilation star with center $c$ and leaves $V$
in expected $O(n \log n)$ time.
\end{theorem}

\begin{proof}
Lemma~\ref{dilation-qcp} shows how to formulate this problem as a
LP-type program with quasiconvex constraints.  Section
\ref{evaluation} provides an $O(n \log n)$ time algorithm to evaluate
the dilation of a star with $n$ leaves; so our decision algorithm can
invoke that algorithm, then compare the result to $\lambda$, to decide
whether $q(\bar{x}) \leq \lambda$ for any $\lambda$.  The resulting
decision algorithm has running time $D(n) \in O(n \log n)$.

A set of input points $P$ may be divided into r subsets, each with
size $\lceil s |P| \rceil$, as follows.  To ensure correctness, every
pairing of points must be mutually present in at least one $P_i$.  So
we partition $P$ into three arbitrary disjoint subsets $Q_1, Q_2,$ and
$Q_3$, each of size $|Q_i| \geq \lfloor |P|/3 \rfloor$, and form three
subsets $P_1=P \setminus Q_1$, $P_2 = P \setminus Q_2$, and $P_3 = P
\setminus Q_3$.  Then any pair of points present in $P$ is also
mutually present in some $P_i$.  Each $P_i$ has size $|P_i| \leq
\frac{2}{3} |P|$, so our approach fits the requirements of Chan's
framework with parameters $r=3$ and $s = \frac{2}{3}$.  Consequently
it may be solved in expected $O(D(n)) = O(n \log n)$ time. 

\hspace{7.55cm} \end{proof}

\section{The constrained optimization \\ problem}

\label{constrained}

Finally we consider the following problem: given a set $V
\subset \mathbb{R}^2$ of $n$ points, find a point $c \in V$ such that the
star with center $c$ and leaves $V \setminus \{c\}$ has minimal
dilation.

At a high level, our algorithm is as follows:\\

\noindent CONSTRAINED-DILATION$(V)$: 
\begin{enumerate}
\item Let $C \leftarrow V$
\item Repeat
\begin{enumerate}
\item Let $c$ be a random point from $C$.
\item Compute $\Delta_c$, the dilation of the star with center $c$ and
leaves $V \setminus \{c\}$, using the algorithm presented in
Section~\ref{evaluation}.
\item \label{compute-region-step} Compute the region $R$ which may
contain a center $c'$ that admits a dilation $\Delta_{c'} < \Delta_c$.
\item $C \leftarrow C \cap R$.
\item If $C=\emptyset$, return $c$.
\end{enumerate}
\end{enumerate}

In each iteration of the loop, any $a \in C$ will be removed with
probability 1/2; so the expected number of iterations is $\log n$.
The remainder of this section will describe a representation of $R$
that may be constructed and queried efficiently.  We will present a
simple data structure that can be constructed in $O(n \, 2^{\alpha(n)}
\log n)$ time and answer membership queries in $O(\log n)$ time, which
makes the overall expected running time of $CONSTRAINED-DILATION$ $O(n
\, 2^{\alpha(n)} \log^2 n)$.

The region $R$ is the intersection of a number of ellipses.  We will
use an argument similar to that of Section~\ref{evaluation} to show
that $R$ may be accurately represented by $O(n)$ ellipses, which may
be identified in $O(n \log n)$ time.  Our algorithm needs to be able
to decide whether given points lie inside, or outside, the
intersection of those $O(n)$ ellipses.  We will show that this
intersection can be described as a sequence of $O(n \,
2^{\alpha(n)})$ arcs.  A straightforward divide-and-conquer algorithm
can construct this sequence in $O(n \, 2^{\alpha(n)} \log n)$ time,
and binary searches in the sequence can answer membership queries in
$O(\log n)$ time.

\begin{lemma} 
\label{r-is-intersection}
For any quantity $\Delta_c$ and $V \subset \mathbb{R}^2$, the region
$R = \{r \enspace | \enspace \Delta_r < \Delta_c \}$ is defined by the
intersection of ellipses.
\end{lemma} 

\begin{figure}
\centering \scalebox{.7}{\includegraphics[bb=1.5in 6.5in 4in
9.7in]{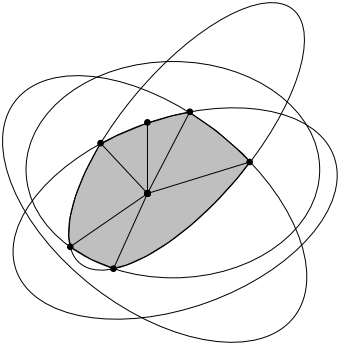}}
\caption{Arcs defining an intersection of ellipses.}
\label{fig:arcs}
\end{figure}

\begin{proof}
The proof of Lemma~\ref{dilation-qcp} contains a description for how
to form a constraint $f_{i,j}$ for each pair of points in $V$.  As
discussed in that proof, the level set $f_{i,j}^{\leq \lambda}$
defines the region of $\mathbb{R}^d$ for which $\Delta_c(v_i,v_j) \leq
\lambda$.  So the region $R$ is the interior of the intersection of
these ellipses; formally,
\[ R = {\rm int} \bigcap_{i \neq j} f_{i,j}^{\leq \Delta_c} \]
(see Figure~\ref{fig:arcs}). 

\hspace{7.55cm} \end{proof}

\begin{lemma}
\label{r-boundary-high}
If $\Delta_c \geq \Gamma$ for a constant $\Gamma>3$, then the region $R$ is the
intersection of $O(n)$ ellipses, which may be identified in $O(n \log
n)$ time.
\end{lemma}

\begin{proof}
Let $a$ and $b$ be the foci of an ellipse contributing to the boundary
of $R$ and containing the boundary point $c'$; then $\Delta_c =
\Delta_{c'} = \Delta_{c'}(a,b)$.  By Lemma~\ref{high-n-points}, $b$
must be one of the $k$ nearest neighbors of $a$.  Therefore, the only
ellipses that contribute to the boundary of $R$ are those found by
reporting the $k$ nearest neighbors of each input
point. 

\hspace{7.55cm} \end{proof}

\begin{lemma}
\label{r-boundary-low}
If $\Delta_c \leq \Gamma$ for a constant $\Gamma$, then the region $R$ is the
intersection of $O(n)$ ellipses, which may be identified in $O(n \log
n)$ expected time.
\end{lemma}

\begin{proof}
As in Lemma~\ref{r-boundary-high}, we will use the algorithm presented
in Section~\ref{evaluation} to identify the relevant ellipses.
However the algorithm presented in Subsection~\ref{low} expects a
center $c$ as input, and we are now computing $c$ as output.  We
overcome this obstacle by computing $c_{OPT}$, the solution to the
unconstrained optimization problem discussed in Section
\ref{unconstrained}.  We then use the approach of
Subsection~\ref{low}, simulating the annuli centered on $c$ with the
annuli centered on $c_{OPT}$.  We will argue that for any $a \in V$ we
can use techniques similar to those in Subsection~\ref{low} to
identify $O(1)$ points, such that $\hat{a}$ is one of those points.
Let $x = |c c_{OPT}|$; then $\hat{a}$ may fall into one of three
categories.  It may be roughly $x$ away from $c_{OPT}$, or closer, or
farther.

Let $a \in V$, and suppose $\rho^{-1} x \leq |a c_{OPT}| \leq \rho x$.
Then $a$ can only lie in one of two annuli centered on $c_{OPT}$,
i.e. $A_{c_{OPT},i}$ or $A_{c_{OPT},i+1}$, for $i=\lfloor \log_{\rho}
x \rfloor$; so for any $a$ we may pessimistically include all the
$O(1)$ points in those annuli in the list of $\hat{a}$ candidates.

Suppose $|a c_{OPT}| > x$; then shifting the annuli's center from $c$
to $c_{OPT}$ will cause $a$ to cross at most one annulus' border.  So
if $a \in A_{c,i}$ then $a \in A_{c_{OPT},i-1} \cup A_{c_{OPT},i} \cup
A_{c_{OPT},i+1}$.  So for these points we include the points within
$2l$ ranks of $a$ in $V_S$, rather than $l$ ranks as in
Subsection~\ref{low}.

Finally we consider points $a$ such that $|a c_{OPT}| < x$.  By the
assumption that $\Delta_{c_{OPT}} \leq \Gamma$ there can only be
$O(1)$ such points; so we can pessimistically include all $O(1)$ such
points.

So for each $a \in V$, we form the ellipses corresponding to $a$
paired with the points in the two annuli whose radii are close to $x$;
the points appearing within $2l$ ranks of $a$ in the sorted sequence
$V_S$; and the $O(1)$ points within $x$ of $c_{OPT}$.

\hspace{7.55cm} \end{proof}

\begin{definition}
An \emph{arc ring} is a pair $(r,S)$ describing the convex region
inside the intersection of ellipses.  $r$ is an arbitrary point inside
the region.  $S=\langle (e_1, \theta_1), (e_2, \theta_2), \ldots
\rangle$ is a sequence of arcs, with each element $(e_i,\theta_i)$
describing a range of angles $\theta_{i-1} \leq \theta \leq
\theta_{i}$ about $r$, for which the ellipse $e_i$ defines the
boundary of the region (where $\theta_{0} \equiv \theta_{|S|}$).  We
require that the angle boundaries $\theta_i$ appear in ascending
order, $\theta_1=0$, and $\theta_{|S|}=2\pi$.
\end{definition}

\begin{lemma}
There exists an arc ring describing any $R$ using $O(n \,
2^{\alpha(n)})$ arcs.
\end{lemma}

\begin{proof}

Lemmas~\ref{r-boundary-high} and~\ref{r-boundary-low} show that for
any value of $\Delta_c$, $R$ may be described as the intersection of
$O(n)$ ellipses.  As shown in \cite{235229}, a description of the
intersection of $O(n)$ ellipses forms a Davenport-Schinzel sequence of
order four; such a sequence has length $O(n \, 2^{\alpha(n)})$.
Hence $|S| \in O(n \, 2^{\alpha(n)})$. 

\hspace{7.55cm} \end{proof}

\begin{lemma} 
An arc ring $E$ describing $R$ may be generated in $O(n \,
2^{\alpha(n)} \log n)$ time and queried in $O(\log n)$ time.
\end{lemma}

\begin{proof}

$E$ may be formed in $O(n \, 2^{\alpha(n)} \log n)$ time using a simple
divide-and-conquer algorithm resembling merge sort.  First partition
the set of ellipses into two subsets of equal size, and compute the
intersections $S_1$ and $S_2$ of the subsequences recursively.  Then
merge $S_1$ and $S_2$, as in merge sort, by comparing the least
elements of each subsequence, $s_1$ and $s_2$ respectively, computing
an element to append to $S=S_1 \cup S_2$, and modifying or deleting
$s_1$ and $s_2$.  This may be done by partitioning the element with a
greater angular interval into two elements, so that $s_1$ and $s_2$
cover exactly the same interval.  If $s_1$ and $s_2$ do not intersect
over that interval, then append whichever element is closer to $r$ to
$S$; otherwise, append two elements corresponding to the closest
ellipse on either side of the intersection point.  The running time of
this algorithm is given by the recurrence
\[ T(|S|) = 2 T(|S|/2) + O(|S|) \enspace , \]
which solves to $T(n) \in O(|S| \log |S|)$; so the running time is
$O(n \, 2^{\alpha(n)} \log (n \, 2^{\alpha(n)})) = O(n \, 2^{\alpha(n)}
\log n)$.  Once $E$ has been constructed, it is possible to determine
whether a point $a$ lies in $R$ by computing the angle between $r$ and
$a$, then performing a binary search in $S$ to find the pertinent arc,
and computing whether $a$ lies inside or outside that arc.  This
operation takes $O(\log n)$ time. 

\hspace{7.55cm} \end{proof}

\begin{theorem} The optimal center $c \in V$ may be found in $O(n \, 2^{\alpha(n)} \log^2 n)$ expected time.
\end{theorem}

\begin{proof}

The expected number of iterations of the outer loop of
$CONSTRAINED-DILATION$ is $\log n$.  The running time of each
iteration is dominated by the construction of $R$, which can be
achieved in $O(n \, 2^{\alpha(n)} \log n)$ time.  Hence the algorithm
runs in $O(n \, 2^{\alpha(n)} \log^2 n)$ expected time. \hspace{1cm}

\hspace{7.55cm} \end{proof}

\bibliographystyle{abbrv}
\bibliography{min-dilation-star}

\begin{thebibliography}{10}

\bibitem{akks-cdpc-02}
P.~K. Agarwal, R.~Klein, C.~Knauer, and M.~Sharir.
\newblock Computing the detour of polygonal curves.
\newblock Technical Report B 02-03, Freie Universit{\"a}t Berlin, Fachbereich
  Mathematik und Informatik, 2002.

\bibitem{AmeBerEpp-Algs-99}
A.~B. Amenta, M.~W. Bern, and D.~Eppstein.
\newblock {Optimal point placement for mesh smoothing}.
\newblock {\em J. Algorithms}, 30(2):302--322, February 1999.
\newblock Special issue for 8th SODA.

\bibitem{982853}
T.~M. Chan.
\newblock An optimal randomized algorithm for maximum {Tukey} depth.
\newblock In {\em Proceedings of the fifteenth annual ACM-SIAM symposium on
  Discrete algorithms}, pages 430--436. Society for Industrial and Applied
  Mathematics, 2004.

\bibitem{location-analysis}
Z.~Drezner and H.~W. Hamacher.
\newblock {\em Facility Location: Applications and Theory}.
\newblock Springer-Verlag, 2001.

\bibitem{Epp-HCG-00}
D.~Eppstein.
\newblock {Spanning trees and spanners}.
\newblock In J.-R. Sack and J.~Urrutia, editors, {\em Handbook of Computational
  Geometry}, chapter~9, pages 425--461. Elsevier, 2000.

\bibitem{eppstein-quasiconvex}
D.~Eppstein.
\newblock Quasiconvex programming.
\newblock \emph{ACM Computing Research Repository}, cs.CG/0412046, 2004.

\bibitem{696169}
S.~Langerman, P.~Morin, and M.~A. Soss.
\newblock Computing the maximum detour and spanning ratio of planar paths,
  trees, and cycles.
\newblock In {\em Proceedings of the 19th Annual Symposium on Theoretical
  Aspects of Computer Science}, pages 250--261. Springer-Verlag, 2002.

\bibitem{586967}
G.~Narasimhan and M.~Smid.
\newblock Approximating the stretch factor of euclidean graphs.
\newblock {\em SIAM J. Comput.}, 30(3):978--989, 2000.

\bibitem{235229}
M.~Sharir and P.~K. Agarwal.
\newblock {\em Davenport-Schinzel sequences and their geometric applications}.
\newblock Cambridge University Press, 1996.

\bibitem{70532}
P.~M. Vaidya.
\newblock An {$O(n \log n)$} algorithm for the all-nearest-neighbors problem.
\newblock {\em Discrete Comput. Geom.}, 4(2):101--115, 1989.

\end{thebibliography}

\end{document}